\documentclass[12pt]{JHEP3}
\input epsf.tex
\epsfclipon

\usepackage{color}
\let\normalcolor\relax
\usepackage{epsfig}
\usepackage{epstopdf}
\usepackage{epsf}
\input{epsf.sty}
\usepackage{graphicx,amsmath,amssymb}
\usepackage{epsfig,multicol}
\usepackage{multirow}
\allowdisplaybreaks

\usepackage{bbm,bm,amsmath,amssymb}

\usepackage[normalem]{ulem}

\usepackage{comment}

\def\figs/B{B}
\def\ba{\begin{eqnarray}}
\def\ea{\end{eqnarray}}
\def\bg{\begin{eqnarray}}
\def\nd{\end{eqnarray}}

\def\log{{\rm log}}
\def\ln{{\rm log}}

\title{On Bulk Viscosity at Weak and Strong\\
't Hooft Couplings}
\author{Alina Czajka$^{1, 2}$, Keshav Dasgupta$^{1}$, Charles Gale$^{1}$, Sangyong Jeon$^{1}$,  
	Aalok Misra$^{3}$, Michael Richard$^{4}$, Karunava Sil$^{5}$\\
	\vskip.03in
	${}^1$ Department of Physics, McGill University \\
	~~3600 rue University, Montr\'{e}al, Qu\'{e}bec, Canada H3A 2T8\\
	${}^2$ Institute of Physics, Jan Kochanowski University\\
	~~Swietokrzyska 15 street, 25-406 Kielce,  Poland\\
	${}^3$  Department of Physics, Indian Institute of Technology Roorkee\\
	~~Uttarakhand-247667, India\\
	${}^4$ John Abbott College, 21275 Lakeshore Dr\\
	~~ Sainte-Anne-de-Bellevue, Qu\'{e}bec, Canada H9X 3L9\\
	${}^5$ Department of Physics, Indian Institute of Technology Ropar\\
	~~Nangal Road, Rupnagar, Punjab 140001, India\\
	{\tt aczajka, keshav, gale, jeon@hep.physics.mcgill.ca}
	~~{\tt michael.richard@mail.mcgill.ca, aalokfph@iitr.ac.in, karunavasil@gmail.com}}

\date{\today}
\maketitle

\abstract{Bulk viscosity is an important transport coefficient that exists in the hydrodynamical limit only when the underlying theory is non-conformal. One example being thermal QCD with large number of colors. We study bulk viscosity in such a theory at low energies and at weak and strong 't Hooft couplings when the temperature is above the deconfinement temperature.  The weak coupling analysis is based on Boltzmann equation from kinetic theory whereas the strong coupling analysis uses non-conformal holographic techniques from string and M-theories. Using these, many properties associated with bulk viscosity may be explicitly derived. This is a shortened companion paper that summarizes some of the results of our longer paper \cite{Bulk2} .}

\begin{document}

\section{Introduction}	

The theory of the strong nuclear interaction is Quantum Chromodynamics (QCD), a nonlinear formalism whose degrees of freedom are quarks and gluons is one of the cornerstones of contemporary subatomic physics \cite{Skands:2012ts} as it successfully describes a wealth of experimental data. However, the extent of our ability to interpret observables in terms of elements of QCD extends only as far as our ability to do controlled calculations. Because of one of the consequences of the nonlinearity of QCD -- asymptotic freedom -- perturbative calculations are possible at large momentum transfer, where the strong coupling constant $\alpha$ is small. As the energy scale becomes softer, the coupling rises and this renders perturbation theory inapplicable. In such a regime, numerical solutions of QCD on  a discretized space-time lattice are possible but still suffer from important limitations. For example, regions where the baryonic density is non-zero require special treatment when applying the Monte Carlo techniques inherent to lattice studies~\cite{Ratti:2018ksb}. In addition to those complications, this numerical  approach is of little use in studies of QCD out of equilibrium, which is the main focus of this paper. 	

In general, the departure from equilibrium can be quantified by the magnitude of the transport coefficients associated with the relevant interaction \cite{LL_1959}. It turns out that the modeling of relativistic nuclear collisions at the energies of RHIC and of the LHC is very efficiently done via relativistic fluid dynamics, and several studies have shown that the collective behaviour of measured hadrons is very sensitive to the value of the shear viscosity of QCD \cite{Gale:2013da}. More recently, it has also become clear that the bulk viscosity of QCD also has an important role to play in the physics of strongly interacting matter in extreme conditions of temperature and density \cite{Ryu:2015vwa}. On the other hand, first bulk viscosity calculations that used systematic methods of relativistic quantum field theory were done in~\cite{Jeon:1994if} and later in~\cite{Benincasa:2005iv}, where a gauge theory was investigated. The importance of bulk viscosity for the nuclear matter dynamics at experimentally achievable energies sets the context in which a recent study of the weak and strong coupling limits of strongly interacting matter were explored, and the evolution of the coefficient of bulk viscosity was examined in this continuum \cite{Bulk2}. It is the purpose of this work to provide a primer of our longer work; the interested reader should find both results and details in these two companion papers. 

The organization of our article is as follows:  The next two sections discuss the calculation of the bulk viscosity of QCD at weak and intermediate couplings, using the techniques of finite temperature field theory. The rest of the paper is devoted to analyses relying on the methods of gauge-string duality. We then discuss and conclude.

\section{Bulk viscosity at weak 't~Hooft coupling \label{weak-kinetic}}

Bulk viscosity $\zeta$ is a transport coefficient characterizing non-conformal systems being out of thermal equilibrium. It determines the deviation from the equilibrium pressure $P$ of expanding or contracting system $\mathcal{P}=P - \zeta \nabla \cdot u$ with $\nabla \cdot u$ - the expansion parameter.

The most efficient framework to compute the coefficient at weak coupling regime is provided by kinetic theory. The calculation was done in Ref.~\cite{Arnold:2006fz} for gauge coupling $g_{YM}$ and can be matched to the 't Hooft coupling $\lambda=g_{YM}^2 M$, where $M \to \infty$ is the number of colors, with not much effort. In such a limit quarks are suppressed by at least a factor of $1/M$ in favor of gluon contributions so it is enough to consider pure gluodynamics for the leading order of bulk viscosity evaluation. Then one needs to solve the Boltzmann equation for the gluon distribution function $f({\bf p},{\bf x},t)$ of the form:
\begin{eqnarray}
\label{Bol-eq}
(\partial_t + {\bf v} \cdot \nabla_{\bf x}) f({\bf p},{\bf x},t) = - \mathcal{C} [f].
\end{eqnarray}
The out-of-equilibrium distribution function can be divided as follows: $f= f_{\rm eq} + f_1$, where $f_{\rm eq}({\bf p},{\bf x},t) = (e^{\beta(t) \gamma_u (E_p (x)-{\bf p}\cdot {\bf u}({\bf x}))} - 1)^{-1}$, with $\gamma_u=(1-u^2)^{-1/2}$. Thus $f_{\rm eq}$ contains space-time dependent quantities: the inverse of temperature $\beta(t)=1/T(t)$ and energy $E_p(x)=\sqrt{{\bf p}^2 + m^2_{\rm th}(x)}$, where $m^2_{\rm th}(x)$ is the thermal mass. $f_1$ is the nonequilibrium correction, which captures both the action of
hydrodynamic forces and the effect of the $x$ dependent quantities.
The collision term $\mathcal{C} [f]$ contains contributions from both the number conserving $gg \to gg$ scatterings and the number changing 
$g\to gg$ splittings and its explicit form is shown in \cite{Arnold:2002zm}.

At the linearized order, the left-hand side of the Boltzmann equation (\ref{Bol-eq}) equals
$ -\beta^2(t) S({\bf p}) \nabla \cdot {\bf u}({\bf x})$ when $\beta(t)=\beta$ and  ${\bf u}({\bf x})=0$. We have also defined
$S({\bf p}) = -T q({\bf p}) f_0(E_p)(1+f_0(E_p))$ with $f_0$ being the Bose-Einstein distribution function $(e^{\beta E_p } - 1)^{-1}$ and the quantity $q({\bf p})$, responsible for the bulk viscosity emergence, is of the form:
\begin{eqnarray}
\label{q1}
q({\bf p}) =\left(\frac{1}{3} -  c_s^2\right) \left[|{\bf p}|- \frac{4\pi^2}{5}
\frac{T^2}{|{\bf p}|}\right].
\end{eqnarray}
For the speed of sound, computed as $c_s^2=\partial P/\partial \epsilon$, one obtains the relation 
$\frac{1}{3} - c_s^2= -\frac{5}{72\pi^2}M \beta_\lambda =
\frac{55}{3456\pi^4} \lambda^2$, where $\beta_\lambda$ is the Callan-Symanzik beta function determining the running of the coupling with the energy scale. Once the left-hand side of the Boltzmann equation is given, one is able to establish the correction to the distribution function $f_1=\beta^2 f_0(f_0+1) \chi \nabla \cdot {\bf u}$, which fixes the form of the collision kernel. The Boltzmann equation can be then expressed as $S({\bf p}) = [\mathcal{C} \chi] ({\bf p})$. Bulk viscosity can be then computed as
$ \zeta = \tilde S_{m} \tilde C^{-1}_{mn} \tilde S_{n} $,
with $\tilde C_{mn} = 2M^2 \int_p \phi_m(p)
[\mathcal{C}\phi_n] (p)$, the column vector is $\tilde S_{m}=2M^2
\int_p \phi_m(p) S(p)$, and the basis functions are $\phi_m(p) = p^m
T^{K-m-1}/(T+p)^{K-2}$ with $m=1,...,K$, for details see Refs.~\cite{Arnold:2000dr,Arnold:2003zc}. Given that, the Boltzmann equation can be solved numerically relying on the variational method. The bulk viscosity to entropy density ratio $\zeta/s$ and the bulk viscosity to shear viscosity ratio $\zeta/\eta$ are then found as the following functions of the parameter $(1/3 - c_s^2)$:
\begin{eqnarray}
\label{bulk-s}
\frac{\zeta}{s} \propto \frac{\lambda^2}{\ln(b_o/\lambda)} \propto
\frac{\left(1/3-c_s^2\right)}{\ln(b_o/\lambda)} , ~~~
\label{sellroth}
{\rm and} ~~~\frac{\zeta}{\eta} \propto \left(1/3-c_s^2\right)^2,
\end{eqnarray}
where $b_o$ is some numerical factor.

Solving the Boltzmann equation is a very efficient way for computing transport coefficients and the truly fundamental prescription is given by the respective Kubo formulas. The one for the bulk viscosity reads:
\begin{eqnarray}
\label{bulk-zeta}
\zeta = \frac{1}{2}\lim_{\omega \to 0}\lim_{{\bf k}\to 0}
\frac{\rho_{PP}(\omega, {\bf k})}{\omega} = \lim_{\omega \to 0}\lim_{{\bf k}\to 0}
\frac{{\text{Im}} G^{PP}_R(\omega, {\bf k})}{\omega} ,
\end{eqnarray}
where $\omega$ is the frequency of the hydrodynamic mode and $\rho_{PP}(\omega, {\bf k})$ is the spectral function of the pressure-pressure correlation function $G^{PP}_R$.
In the weakly coupled regime the spectral function can be computed diagrammatically. While a full quantitative analysis is highly nontrivial, we provide here qualitative analysis on scattering processes and corresponding diagrams, which dominate transport phenomena. As discussed for QED transport coefficients in~\cite{Gagnon:2006hi,Gagnon:2007qt}, the equivalence of kinetic theory to the diagrammatic approach can be established when one shows that only planar diagrams govern the collision kernel of the Boltzmann equation. We therefore present topological structures of planar ladder diagrams of the SU$(M)$ theory which contribute to the Boltzmann equation and perform power counting of the corresponding processes. This provides a solid argument to justify the validity of the Boltzmann equation. 

\begin{figure}[h]
	\centering
	\begin{tabular}{c}
		\includegraphics[width=0.5\textwidth]{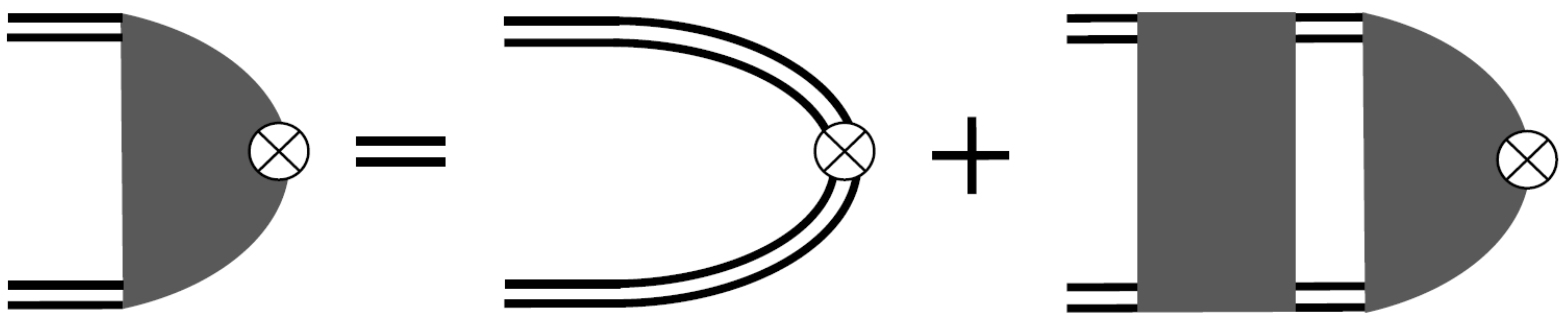}
	\end{tabular}
	\caption{The integral equation needed for bulk viscosity evaluation in the SU($M$) theory.}
	\label{series}
\end{figure}

\begin{figure}[h]
	\centering
	\begin{tabular}{c}
		\includegraphics[width=0.8\textwidth]{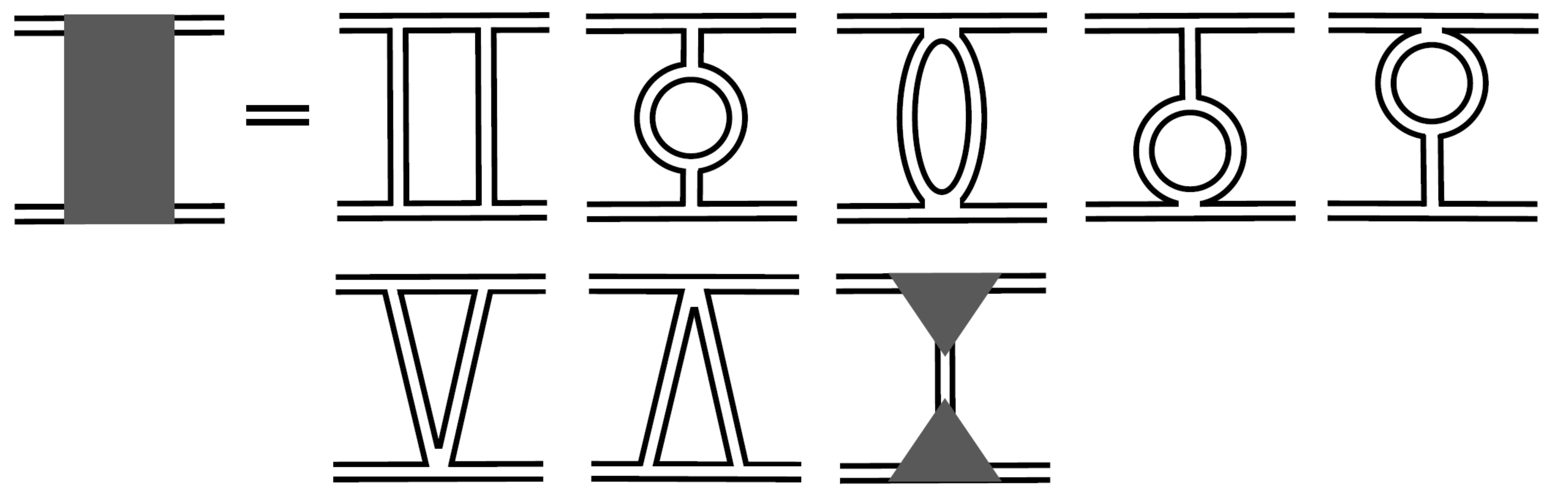}
	\end{tabular}
	\caption{The schematic form of the kernel of the integral equation with possible topological rung
		insertions.}
	\label{series-k}
\end{figure}

\begin{figure}[h]
	\centering
	\begin{tabular}{c}
		\includegraphics[width=0.8\textwidth]{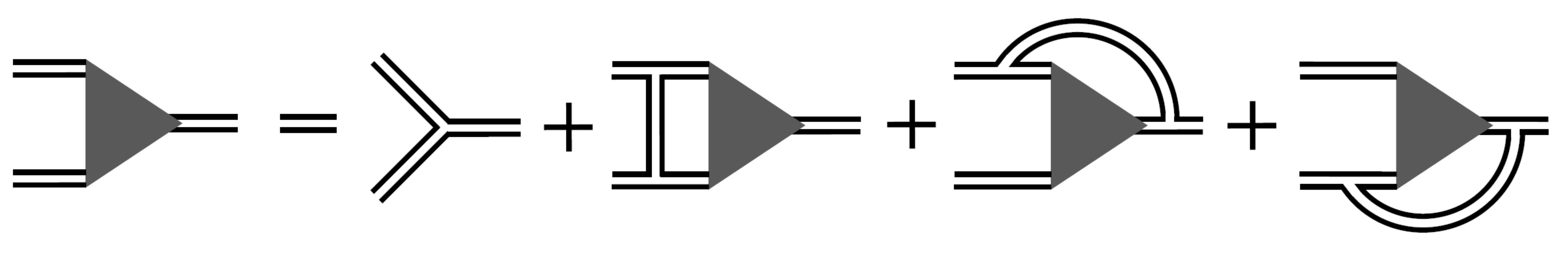}
	\end{tabular}
	\caption{Integral equation for the effective vertex characteristic of the
		collinear splittings.}
	\label{series-v}
\end{figure}

The analysis of diagrams contributing to the bulk viscosity spectral function is most conveniently performed in the $(r,a)$ basis of real time quantum field theory, where the retarded propagator $G_{ra}$, advanced one $G_{ar}$ and the auto-correlation function $G_{rr}=(1+2n_B)(G_{ra}-G_{ar})$, with $n_B$ - the Bose-Einstein distribution function, determine the structure of diagrams. The dominant contribution to the leading order bulk viscosity is established by the terms $G_{ra}(p)G_{ar}(p)$, which have a singularity introduced by the pinching poles or nearly pinching poles. The singularities are cured when all propagators are dressed with the self-energy $\Pi(p)$. The real part of the self-energy constitutes the thermal mass $m_{\rm th}^2 \propto \lambda T^2$ and the imaginary part of the order of $O(\lambda^2 T^2)$, obtained at the two-loop order, is related to the thermal width $\Gamma_p$, which governs the location of the poles. The typical size of the bulk viscosity is then of the order of $(1/3-c_s^2)^2/\Gamma_p$. The bulk viscosity is fully determined when all diagrams of a given size are resummed. Any rung of the order $O(\lambda^2)$ can therefore be inserted in the spectral function loop without changing its parametric form since the contribution from the rung cancels the one from the newly emergent pair of the retarded and advanced propagators. The resummation of all possibilities is given in the form of the integral equation shown schematically in Fig.~\ref{series}, where we adopted the 't~Hooft double line notation to represent gluon propagators. All distinct topological rungs determine the kernel of the integral equation and they are shown schematically in Fig.~\ref{series-k}. The structures of rungs are found from the cuts of the two-loop self-energy when all possible constraints are taken into account: the Ward identity, power counting and allowed kinematic regions \cite{Gagnon:2006hi,Gagnon:2007qt}. The detailed power counting of each rung and the role soft physics plays there is presented in our paper \cite{Bulk2}. Here we only emphasize that the diagrams with the pinching pole singularities, when the collinear singularities are absent, represent number conserving $2 \to 2$ scatterings. When the collinear singularities are included, the emergent diagrams correspond to $1+N \to 2+N$ collinear gluon splittings. In the latter case the role of soft physics is crucial as the splitting of a hard quasiparticle is possible when it undergoes a soft interaction with the thermal medium. In such a process it is important to notice the role of $G_{rr}$ propagator, which contains the phase space density and thus represents the soft particle exchange. Infinitely many $G_{rr}$ insertions are possible in the leading order analysis and all of them need to be resummed capturing Landau-Pomeranchuk-Migdal (LPM) effect. This leads to the integral equation for the effective vertex shown in Fig.~\ref{series-v}. The solution of the later one needs to be inserted to the kernel of the integral equation for the spectral function. The integral equation for the spectral function can be thought as the diagrammatic representation of the Boltzmann equation used for the bulk (and shear) viscosity evaluation.

\section{Bulk viscosity at intermediate coupling \label{int}}

In the regime near the phase transition, the coupling is strong enough that perturbative methods are not applicable. The microscopic approach that can be used here relies on the Euclidean Lattice QCD (LQCD).  A few attempts have been made to extract bulk viscosity in this domain from the lattice data with the help of the QCD sum rules in particular. To make use of the LQCD results one shall start with the Kubo formula given by Eq. (\ref{bulk-zeta}). Note, however, that numerical factor convention and the operator, for which the correlation function is studied, can be slightly different. In Refs.~\cite{Karsch:2007jc,Romatschke:2009ng,Meyer:2010ii} the following versions of the QCD sum rule were studied, respectively:
\begin{eqnarray}
&&\int_0^\infty {d\omega\over \pi }\, {\rho_{\Theta\Theta}(\omega, \mathbf{0})\over\omega}
=
\left( T{\partial\over \partial T} - 4\right)\langle \Theta_G \rangle_T
+ (\hbox{quark contribution}),
\label{eq:KKT} \\
&&\int_0^\infty {d\omega\over \pi} {\delta\rho_{\Theta\Theta}(\omega, \mathbf{0})\over \omega}
=
\left( 3s {\partial\over \partial s} - 4\right)\left(\epsilon - 3P\right),
\label{eq:SR} \\
&&\int_{0}^\infty {d\omega\over \pi}
{\delta\rho_{*}(\omega,\mathbf{0}) \over\omega}
=
3(1 - 3c_s^2)(\epsilon+P) - 4(\epsilon - 3P),
\label{eq:Meyer}
\end{eqnarray}
where $\Theta_G$ is the gluon contribution to the trace of the stress-energy tensor
and $\rho_{\Theta\Theta}$ is the spectral density for the $\Theta\Theta$ correlation 
function with $\hat\Theta=\hat T^\mu_\mu=\hat T^{00}-3\hat P $. The trace average is $\langle \Theta_G \rangle_T = \left(\epsilon - 3P\right) + \langle \Theta_G \rangle_0$, where $\langle \Theta_G \rangle_0$ is the vacuum contribution. For the purpose of this analysis the quark contribution in Eq.~(\ref{eq:KKT}) is not essential. In Eq.~(\ref{eq:SR}) the deviation from the vacuum spectral density at finite temperature $\delta\rho_{\Theta\Theta} =\rho_T -\rho_0 $ is used and $\delta\rho_{*}$ in Eq.~(\ref{eq:Meyer}) is defined for the operator $\hat\Theta_* = 
\hat T^{\mu}_{\mu} - (1 - 3c_s^2)\hat T^{00}$. While the difference between Eqs.~(\ref{eq:KKT}) and (\ref{eq:SR}) comes from the non-commutability of the limits $\omega\to 0$ and ${\bf k}\to 0$, which is discussed in Ref.~\cite{Romatschke:2009ng}, in Eq.~(\ref{eq:Meyer}) these limits commute. 

If one can relate left-hand side of any equation among (\ref{eq:KKT}-\ref{eq:Meyer}) to the bulk viscosity, then LQCD findings can be applied to the right-hand sides of the equations and the coefficient could be extracted. In Ref.~\cite{Karsch:2007jc} a single Lorentzian ansatz for the spectral density was proposed and the bulk viscosity was then extracted. The consequences of the sum rule and the form of the ansatz were then discussed in Ref. \cite{Romatschke:2009ng,Meyer:2010ii,Moore:2008ws}. It then turned out that the ansatz does not include the contribution from frequencies higher than $\omega_0$, which is negative and cancel the low frequency part. The other problem is that the ansatz makes the left-hand side of Eq.~(\ref{eq:Meyer}) positive while the right-hand side is negative. There are some suggestions that the presence of glueballs may cause the difference. All these arguments lead to the conclusion that the constraints coming from the sum rule are too weak to extract bulk viscosity reliably.

On the other hand, the LQCD calculations can provide information on the static properties of the medium. These could be in principle analytically continued to the real-time space where the bulk as well as the shear viscosities could be evaluated. This procedure, however, leads to very large uncertainties and one can only extract that near the critical temperature $\delta\rho_{*}(\omega, \mathbf{0})/ \omega$ is noticeably enhanced at $\omega=0$ and the order of magnitude of $\zeta/s$ is $O(10^{-2}) - O(10^{-1})$. 

\section{Bulk viscosity at weak string and strong 't Hooft couplings  \label{weak}}

Bulk viscosity at strong 't Hooft coupling maybe studied from either weak string coupling or strong string coupling. All we need is that the number of colors may be large enough so that the 't Hooft coupling $\lambda$ can be very large.
For any of these two cases, string theory is necessary because the strong 't Hooft coupling regime can neither be  accessed by the kinetic theory nor by LQCD. The question is why is this the case? 

The answer lies in the fact that, at strong 't Hooft coupling, the dynamics of QCD with large number of colors and at low energies can be studied by a dual {\it classical} gravity description. The dual description is not an 
$AdS_5 \times \mathbb{M}_5$  space, but a more complicated space given by a resolved warped-deformed conifold with background fluxes. The resolution parameter is essential to have a sensible UV completion of the corresponding theory. The UV however is not a 
SU($M$) theory as one might have expected, but is a more non-trivial theory given by a product gauge group
SU($N+M$) $\times$ SU($N+M$) which is a CFT. The full UV theory, in the type IIB side,  is first discussed in 
\cite{metrics} and \cite{3regions}, which in turn is based on the IR story originally proposed in \cite{KS} and 
\cite{ouyang}. 

The appearance of $N$ and $M$ colors in the description of the gauge groups at IR and UV deserves some explanation. They are the remnants of the number of D3 and D5-branes in the gauge theory side. At the UV the asymptotic CFT has a walking RG flow and the gauge group gets Higgsed to SU($N+M$) $\times$ SU($N$) at a certain scale. After which cascading starts and the far IR theory is given by a SU($M)$  gauge theory that confines. 
The complete flow from UV to IR has recently been discussed using a T-dual type IIA configuration in 
\cite{UVcom}\footnote{For an alternative bottom-up approach of studying bulk viscosity and other transport coefficients the readers may refer to \cite{Attempt} where a dual background for a gauge theory with UV and IR fixed points has been proposed.}.

The analysis that we present here, which in turn is based on section 4 of \cite{Bulk2}, uses weak string coupling. However there exists another limit with strong $g_s$ (i.e $g_s \sim {\cal O}(1)$), which is discussed in section 5 of 
\cite{Bulk2}. The strong $g_s$ limit was first developed in \cite{MQGP}
using certain sequences of string dualities and M-theory uplift, that still keep us in the supergravity regime, albeit from eleven dimensional point of view. This will be elaborated in section \ref{strong}.

The analysis that we present in section 4 of \cite{Bulk2} uses the fluctuations of the background vielbeins to study bulk viscosity. A study along this direction was previously attempted in \cite{bulky} wherein the focus was mostly to build up a consistent setting on which concrete computations may be performed. In \cite{Bulk2}, the story was completed in full details, and the result we get for the ratio of the bulk viscosity, $\zeta$, and the entropy density, $s$, may be succinctly expressed as: 
\bg\label{bulky2}
{\zeta\over s} = {3\epsilon Y_x(r_h, 0) r_h\over 64}\left[3 + {13r_c\over r_h}\left({r_c^4\over r_h^4} - {16\over 13}\right){Y_x(r_c, 0)\over Y_x(r_h, 0)}\right], \nd
where $r_c$ is the cut-off radius, $r_h$ is the horizon radius that gives us temperature and $Y_x(r, \omega)$ is related to the fluctuation at $\omega = 0$. In fact $Y_x$ is precisely proportional to $p_{0x}$ that appears in the fluctuation analysis in eq. (4.13) of \cite{Bulk2}.  It also turns out that the $Y_x$ fluctuations form the key basis on which the computations of bulk viscosity is based on in section 4 of \cite{Bulk2}. These fluctuations satisfy 
a second order differential equation of the form:
\bg\label{HKP}
 a_{12} {d^2Y_x\over dr^2} + a_{22} {dY_x\over dr} + a_{32} Y_x = a_{42}, \nd
 where the $a_{I2}$ coefficients are given in eq. (4.51) of \cite{Bulk2}, with $a_{42}$ being related to certain sources coming from the fluxes and branes in Regions 2 and 3 of \cite{3regions}. 
 
 The factor of $r_h$, the horizon radius, appearing in \eqref{bulky2} tells us that the ratio of bulk viscosity over the entropy density will be related to the temperature. However the relation between the temperature and $r_h$ is more non-trivial. For the present case this may be written as:
 \bg\label{beirut}
 T = r_h\left(a_1 + \epsilon a_2\right), \nd
 where $\epsilon = {3g_sM^2\over 2\pi N}$ is the non-conformality factor
 and $a_i \equiv a_i(x)$ are parameters that depend on $x \equiv {r_h^2\over r_c^2}$. As discussed in \cite{Bulk2}, $a_1$ can be taken to be a constant, but $a_2$ remains a non-trivial function of $x$ as given in eq. (4.78) of \cite{Bulk2}. 
 All of these parameters are in turn crucial to define the sound speed $c_s^2$ in the following way: 
\bg\label{attic2}
c_s^2 & = & {1\over 3} -{2\epsilon\over 45}\sum_{n = 1}^\infty {x^{2n}\over n(2n-1)}\left(1 + {2x\over a_1} {da_1\over dx}\right)\\
&+& {4x^2\over 9a_1^2}\left[\left({da_1\over dx}\right)^2\left(1 - {2\epsilon a_2\over a_1}\right) + {2\epsilon} {da_1\over dx} {da_2\over dx}\right]
+ {8x\over 9a_1}\left[{da_1\over dx}\left(1 - {\epsilon a_2\over a_1}\right) + \epsilon {da_2\over dx}\right], \nonumber \nd
with $\epsilon$ being the expansion parameter discussed above. Expectedly the sound speed is equal to its conformal value when $\epsilon = 0$, and becomes smaller than ${1\over \sqrt{3}}$ once we switch on non-conformal corrections.
Note that  \eqref{bulky2} and the deviation from conformality in \eqref{beirut}
are both proportional to $\epsilon$ as one might have expected. This means if we want the ratio of \eqref{bulky2} and 
$\eta/s$ $-$ $\eta$ being the shear viscosity $-$ to the first order in $\epsilon$, we can simply take the conformal value of ${\eta \over s} = {1\over 4\pi}$ \cite{KSS}.  Putting everything together, and performing some manipulations, gives us the following bound:
\bg\label{mathadulaiC}
{\zeta\over \eta}  >  {405 d_1\over 16}\left({1\over 3} - c_s^2\right), \nd
where the reader can get all the algebraic details from \cite{Bulk2}. Clearly the bound depends linearly on the deviation of the sound speed from its conformal value. This should be compared to what we had before at weak 't Hooft coupling in \eqref{sellroth}. Our result \eqref{mathadulaiC} is expressed in terms of a parameter $d_1$ that can be shown to have the following range: 
\bg\label{duldulC}
{32\over 405}  \le d_1 < {13b r_c^4\over \left(13b + {640\vert c_2\vert}\right)r_h^4}, \nd
where $b$ and $c_2$ are some constant pieces in $a_1$ and $a_2$ of \eqref{beirut} respectively and $r_c$ is the cut-off. The range of $d_1$ in \eqref{duldulC} is well within the Buchel-bound \cite{Buchel-bound}.

\section{Bulk viscosity at strong string and strong  't Hooft couplings \label{strong}}
	
From the point of view of constructing a holographic dual truly close to thermal QCD-like theories, one would have to consider finite gauge coupling and finte number of colors. From the perspective of gauge-gravity duality, this entails looking at the strong-coupling/non-perturbative limit of string theory - M theory. It is precisely this limit $-$ dubbed as the `MQGP limit' $-$ that was looked at in \cite{MQGP} wherein $g_s\stackrel{<}{\sim}1, N_f \equiv {\cal O}(1), (N, M) \gg1$ such that $\frac{g_s M^2}{N}\ll1$.\footnote{$M$ could also be ${\cal O}(1)$ such that the non-planar diagrams are still suppressed by $1/M$.} 
The M-theory uplift of the type IIB holographic dual of \cite{metrics} was constructed in \cite{MQGP} by working out the SYZ type IIA mirror of \cite{metrics} implemented via a triple T duality along a local special Lagrangian (sLag) $T^3$ $-$ which could be identified with the $T^2$-invariant sLag of \cite{M.Ionel and M.Min-OO (2008)} $-$ in the large-complex structure limit effected by making the base ${\cal B}(r,\theta_1,\theta_2)$ (of a $T^3(\phi_1,\phi_2,\psi)$-fibration over ${\cal B}(r,\theta_1,\theta_2)$) large \cite{MQGP,NPB}. The basic idea then is the following. Consider $N$ D3-branes  oriented along $x^{0, 1, 2, 3}$ at the tip of conifold with $M$ D5-branes, parallel to the D3-branes and wrapping a vanishing $S^2(\theta_2,\phi_2)$. A single T-dual of this along $\psi$ (or the corresponding $T^3$ coordinate) yields $N$ D4-branes going all the way along the $\psi$ circle and M D4-branes straddling a pair of orthogonal NS5-branes. These NS5-branes 
correspond to the vanishing $S^2(\theta_2,\phi_2)$ and the blown-up $S^2(\theta_1,\phi_1)$ with a non-zero resolution parameter $a$. Two further T-dualities along $\phi_i$ and $\phi_2$, in the absence of the straddling D4-branes would convert the two orthogonal NS5-branes into two orthogonal Taub-NUT spaces, and the $N$ D4-branes into $N$ D6-branes. In the presence of the $M$ straddling D4-branes, which are originally fractional three-branes in the type IIB side, would eventually T-dualize also to six-branes. Similarly, in the presence of $N_f$ flavor D7-branes, oriented parallel to the three-branes and wrapping a 4-cycle given by ($r, \psi, \theta_1, \phi_1$), T-dualize to $N_f$
D6-branes wrapping a 3-cycle given by ($r, \theta_1, \phi_2$).\footnote{Both Gauss' law as well as kappa-symmetry can be shown to be satisfied in this setup.}. A further uplift to M-theory will convert  the D6-branes to KK monopoles, which are variants of the Taub-NUT spaces discussed above.  Therefore all the branes have converted to geometry and fluxes, and after the dust settles, one  ends with M-theory on a $G_2$-structure manifold.
Similarly, one may perform identical three T-dualities on the gravity dual on the type IIB side, which is a resolved warped-deformed conifold with fluxes,  to finally land in M-theory on another $G_2$ structure manifold, giving us the MQGP model of  \cite{MQGP,NPB}.  

In this setup, by working near the following choices of the angular coordinates $\theta_1$ and $\theta_2$:
\begin{equation}\label{harper}
\theta_1\sim {\alpha_{\theta_1}\over N^{\frac{1}{5}}}, ~~~~~~~ \theta_2\sim {\alpha_{\theta_2}\over {N^{\frac{3}{10}}}},
\end{equation}
$\alpha_{\theta_{1,2}}$ being ${\cal O}(1)$, we can allow the decoupling of the 
five-dimensional spacetime $\mathbb{M}_5$, oriented along $\left(x^0 = t, x^{1, 2, 3}, u = {r_h\over r}\right)$ from the internal six-dimensional space $\mathbb{M}_6$, oriented along ($\theta_{1,2},\phi_{1,2},\psi,x^{10}$). Also, near (\ref{harper}), explicit $SU(3)$ structures for type IIB and its SYZ type IIA mirror and $G_2$ structure for the M-theory uplift were worked out in \cite{NPB}. 

Using the ideas developed in \cite{klebanov quasinormal} and \cite{EPJC-2}, having integrated out the six angular	directions, up to NLO in $N$ in the MQGP limit of \cite{MQGP}, a gauge-invariant combination $Z_s(u)$ of scalar modes $h_{\mu\nu}$ of (M-theory) metric perturbations   invariant under infinitesimal diffeomorphisms:
\begin{equation}\label{nova}
h_{\mu\nu} ~\rightarrow ~ h_{\mu\nu}+\nabla_{(\mu}\xi_{\nu)},
\end{equation}
was constructed in \cite{Bulk2}. It is further shown in \cite{Bulk2} that $Z_s(u)$ satisfies:
\begin{equation}\label{Z-EOM}
Z_s^{\prime\prime}(u) = m(u) Z_s^\prime(u) + l(u) Z_s(u),
\end{equation}
which for a (non-)trivial bare resolution parameter has the horizon ($u=1$) as a(n) (ir)regular singular point. The detailed expressions for $m(u)$ and $l(u)$ appear in eq. (5.15) and eq. (5.21) respectively of \cite{Bulk2}. Substituting the following ansatz for the dispersion relation:
\begin{eqnarray}
\label{dispersion}
\omega_3 = \left(\frac{1}{\sqrt{3}} + \alpha \frac{g_s M^2}{N}\right)q_3 + \left(-\frac{i}{6} + \beta\frac{g_s M^2}{N}\right)q_3^2,
\end{eqnarray}
 into (\ref{Z-EOM}), and by further making an ansatz: $Z_s(u\sim1)\sim e^{S(u)}$ wiith $\left[S^\prime(u\sim1)\right]^2\gg|S^{\prime\prime}(u\sim1)|$ and demanding that the residue of $S^\prime(u\rightarrow1)$ vanishes\footnote{This  ensures that $u=1$ is not a regular singular point.},  it was shown in \cite{Bulk2} that one obtains the following values of $b, \alpha$ and $\beta$:
 
 {\footnotesize
 \bg\label{nullres}
 b \approx \sqrt{6}, ~~ 
  \alpha = {\sqrt{3}{\cal C}_{21}(1)\over 32\pi}  - {c_1 + c_2\log~r_h\over 6\sqrt{2}}, ~~
\beta = -{{3i}{\cal C}_{21}(1)\over 64} - {i\sqrt{6}\left(c_1 + c_2\log~r_h\right)\over 72}, \nd}
where  ($c_1, c_2$) have been defined in eq. (5.52) and eq. (5.57) of \cite{Bulk2} and 
${\cal C}_{kj}$ appearing above may be defined in the following way:
\begin{equation}\label{jkay85}
{\cal C}_{kj}(u) \equiv 1 + {g_sN_f \over 4\pi} \log \left({\alpha_{\theta_1} \alpha_{\theta_2}\over 4\sqrt{N}}\right) + {3g_s N_f\over 2\pi} \left(k ~\log~{r_h\over u} + {2j -1\over 2}\right), \end{equation}
where ($k, j$) will be integers. Combining everything together, and following the detailed analysis in section 5 of 
\cite{Bulk2}, 
one obtains the following functional form for the sound speed $c_s$ and the attenuation constant $\Gamma$:
\bg\label{csG}
&&c_s \equiv {1\over \sqrt{3}} + {\sqrt{3}\over 32\pi}\left({g_sM^2\over N}\right) {\cal C}_{21}(1) - {g_sM^2\over 6\sqrt{2} N}\left(c_1 + c_2\log~r_h\right) \nonumber\\
&& \Gamma  \equiv {1\over \pi T}\left[{1\over 6} + {3 g_sM^2\over 64\pi N}\left({\cal C}_{21}(1) + {8\sqrt{6}\pi\over 27}\left(c_1 + c_2\log~r_h\right)\right)\right],
\nd
with $T$ being the temperature and ($c_1, c_2$) being the same coefficients that appeared in \eqref{nullres} above. It is easy to see that we naturally reproduce the correct conformal results. 

It was explicitly shown in \cite{Bulk2} that in the absence of the bare resolution parameter, one can not consistently impose Dirichlet boundary condition (at the asymptotic boundary) without allowing non-normalizable modes to propagate. 
Then using results from \cite{MQGP,EPJC-2}, it was shown in eq. (5.50) of  \cite{Bulk2} that ${\eta\over s}$ may also be expressed using the ($c_1, c_2$) coefficients defined above. Looking at the values for ($c_1, c_2$)  one may easily infer that the ratio of the shear viscosity over entropy density is bigger than ${1\over 4\pi}$, i.e  
${\eta\over s} > {1\over 4\pi}$,  and hence the KSS \cite{KSS} bound is not violated. 

To complete the story, what is now required is to find a relation between bulk viscosity, $\zeta$, shear viscosity, $\eta$ and the entropy density $s$. This is where the attenuation constant $\Gamma$ given in \eqref{csG} becomes useful because:
\begin{equation}\label{circle}
{1\over 2sT}\left(\zeta + {4\eta\over 3}\right) = \Gamma.  \end{equation}
The extra factor of $T$ appearing above gets cancelled from the inverse $T$ dependence of $\Gamma$ in \eqref{csG}. Therefore, plugging the functional form for ${\eta\over s}$ from eq. (5.50) in \cite{Bulk2}, one may get the functional form for ${\zeta\over s}$. From this we can easily get: 
\begin{eqnarray}\label{vedpra}
{\zeta\over \eta} & = &  {91\over 5}\left({1\over 3} - c_s^2\right) +  {g_sM^2\over 16 \pi N}\Bigg[{201\over 5}
+ {121\over 20\pi}\left(\log\left(\alpha_{\theta_1}\alpha_{\theta_2}\right) + {603\over 121}\right)g_sN_f \nonumber\\
& +& {2g_sN_f\over \pi} \left(\log~N - {603\over 10}\right) \left(\log~{r_c\over r_h} - {201\over 80}\right) - \sigma_0 g_sN_f\Bigg], \end{eqnarray}
where $\sigma_0 \equiv {201\over 20\pi}\left(\log~4 + {603\over 20}\right) \approx 100.86$. Every
term in \eqref{vedpra} is positive definite,
and the negative piece $\sigma_0g_sN_f$ does not make a difference provided $\log~N~\log~{r_c\over r_h} >> 160$. Thus we get a crisp bound:
\bg\label{goeskalu}
{\zeta\over \eta} ~ > ~  {91\over 5}\left({1\over 3} - c_s^2\right). \nd 

\section{Gauge spectral function at strong coupling and non-zero flavors \label{spectralf}}

In this section we summarize the evaluation at strong coupling of a gauge-fluctuation spectral function in line with the discussions of section \ref{int}. Let us start by the following two observations:

\vskip.1in

\noindent $\bullet$ The ratio  $\frac{\zeta}{s}\equiv{\cal O}\left({g_sM^2\over N}\right)$ and 
$\frac{\eta}{s} = \frac{1}{4\pi} + {\cal O}\left({g_sM^2\over N}\right)$-correction term as found in 
section 5 of \cite{Bulk2} implies that up to 
${\cal O}\left({1\over N}\right)$,  the ratio $\zeta/\eta$ would mimic the ratio $\zeta/s$.

\vskip.1in

\noindent $\bullet$ The  gauge and the metric perturbations may be required to be considered simultaneously as discussed in 
subsection 4.2 of \cite{D.Tong_Cracow_13} (and references therein). This means the correlation of gauge fluctuations,  
$\langle {\cal A}_{x^i}{\cal A}_{x^i}\rangle$ for $i=1, 2, 3$, along the same direction could hence mimic the spirit behind the correlation of the metric perturbations, 
$\langle h_{x^ix^i}h_{x^ix^i}\rangle$,  along the $x^i$ axis relevant to the evaluation of bulk viscosity as \cite{Gubser:2008yx},

\vskip.1in

The above observations
 provide the necessary motivation for the evaluation of the aforementioned gauge-field correlation function (in the hydrodynamical limit using the prescription of \cite{Minkowskian-correlators}) and see if one obtains a linear bound seen earlier. Even if  not be explicitly related to $\zeta/\eta$, we feel the result obtained in this section, in its own right, is sufficiently interesting.   
	
Our starting point  then is the DBI action for $N_f$ $D6$ branes in the fundamental representation given as:
\begin{equation}\label{D6DBI}
S_{D6}=-T_{D6}N_f\int d^{7}\xi~ e^{-\varphi}\sqrt{\det{\left(g+B+F\right)}},
\end{equation}
with $2\pi\alpha^{\prime}=1$. In (\ref{D6DBI}) the worldvolume directions of the $D6$ brane are denoted by the coordinates: 
$\left(t, x^1, x^2, x^3, Z,\theta_2, \varphi_2\right)$, with $\left( t, x^1, x^2, x^3\right)$ as the usual Minkowski coordinate, 
$Z$ as the newly defined dimensionless  radial direction and two angular coordinate ($\theta_2$, $\varphi_2$); $Z$ is related to  $r$ as $r = r_{h}e^{Z}$ and $\varphi_2$ is the local value for the angle $\phi_2$  \cite{meson Yadav+Misra+Sil}. In the above, $\varphi$ denotes the type IIA dilaton which is the triple T-dual version of type IIB dilaton. The pullback metric and the pullback of the NS-NS $B$ field on the worldvolume of the $D6$ brane are denoted as $g$ and $B$ in (\ref{D6DBI}). $F$ is the field strength for a U$(1)$ gauge field $A_{\mu}$, where only $A_{t}$ is assumed to be non-zero. In the gauge $A_{Z}=0$, only $F_{Zt}=-F_{tZ}\neq0$. Combining together the symmetric $g$ field and the anti-symmetric $B$ field as 
$G \equiv g+B$, the DBI action is expanded up to quadratic order in $A$ and thereafter the EOM for $A_t$ is 	found to be:
\begin{equation}\label{EOMAt}
\partial_{Z}\left(e^{-\varphi}\sqrt{-G}~G^{tt}G^{ZZ}\partial_{Z}A_{t}(Z)\right)= 0.
\end{equation}
The solution to (\ref{EOMAt}) for $Z\gg1$ is:
\bg\label{solution-At-EOM}
\langle A_t(Z)\rangle 
 =   \frac{{\bf C} e^{-2 Z}}{12 \alpha_{\theta_1}^4 g_s  {N_f} {r_h}^2 Z}+ {\bf C}_1 + {\cal O}\left(\frac{1}{Z^2}\right),
\nd
where $\langle A_t\rangle$ was used to express the background value to avoid confusion and 
${\bf C}_1$  is a constant. Consider fluctuations about the background value of the gauge field in the following way:
\begin{equation}\label{oohswt}
{A}_{\mu}(x, Z)=\delta^{t}_{\mu}\langle A_{t}(Z) \rangle +{\cal A}_{\mu}(x,Z),
\end{equation}
where the fluctuation ${\cal A}_{\mu}$ only exists along the directions $\mu=$ ($ t, x^1, x^2, x^3$) due to the particular gauge choice and depends only on the radial variable $Z$.
Including the perturbations in the DBI action (\ref{D6DBI}), one obtains:
\begin{equation}
\mathcal{L}=e^{-\varphi}\sqrt{\det\left(g+B+F+{\cal F}\right)},
\end{equation} 
with ${\cal F}$ as the field strength for the gauge field fluctuations. Defining ${\cal G} \equiv g+B+F$ and again expanding the above lagrangian upto quadratic order in the gauge field fluctuation, yields the equation of motion for the gauge field fluctuation:
\begin{equation}\label{EOMtildeA} 
\partial_{\alpha}\Biggl[e^{-\varphi}\sqrt{-{\cal G}}\left({\cal G}^{\mu[\alpha}{\cal G}^{\beta]\gamma}
\partial_{[\gamma}{\cal A}_{\mu]}
-\frac{1}{2}{\cal G}^{[\alpha\beta]}{\cal G}^{\mu\nu}\partial_{[\mu}{\cal A}_{\nu]}
\right)\Bigg] = 0.
\end{equation}
We now go the dual Fourier momentum space and work with gauge-invariant gauge fluctuation variables -- electric field -- $E_{x^1}$ and $E_\beta$ with $\beta = x^2$ or $x^3$, expressed in the following way:
$E_{x^1} \equiv q {\cal A}_t + \omega {\cal A}_{x^1}, ~ E_\beta \equiv E_T = \omega {\cal A}_\beta$ (with a slight abuse of notation wherein the coordinate-space and momentum-space valued gauge fluctuations are denoted by the same ${\cal A}_{t, x^{1,2,3}}$). In the $q=0$-limit, the EOMs for $E_{x^1}$ and $E_T$ coincide. In was shown in \cite{Bulk2} that the same can be rewritten as a Schr\"{o}dinger-like equation: 
\begin{equation}
\label{Schrodie}
\left(\partial^{2}_{Z}+ \mathbb{V}_{E_{T}}\right)\mathbb{E}_{T} = 0,
\end{equation}
with $\mathbb{E}_T$ and $\mathbb{V}_T$ are given in eq (6.22) and (6.30) of \cite{Bulk2}. 
One notes that the horizon -- $Z=0$ -- is a regular singular point of (\ref{Schrodie}) implying one could make the ansatz: $\mathbb{E}_T = Z^{{}^{{1\over 2} - i\mathbb{I}}}~\mathbb{F}_T(Z)$, $\mathbb{F}_T(Z)$ being analytic for all $Z$'s.  Substituting the aforementioned ansatz into (\ref{Schrodie}), and solving as shown in \cite{Bulk2}, one obtains for $Z\gg1$:
\begin{equation}
\label{ET_UV}
\mathbb{E}_T(Z) =  Z^{{}^{{1\over 2} - i \mathbb{I}}} 
\left[ {\bf C}_+ {\rm exp}\left(-{iZ\over \sqrt{\bf A}}\right) + {\bf C}_- {\rm exp}\left({iZ\over \sqrt{\bf A}}\right)\right], 
\end{equation}
where $C_+$ and $C_-$ are two integration constants.
As shown in \cite{Bulk2}, one obtains the following on-shell action for the $x^1$ piece of the fluctuation: 
\begin{equation}\label{asidana}
\mathbb{S}_4^{(1)}= - {\Omega_2 T_{D6} \over  2 \omega^2}\int d^4x \Biggl[e^{-\varphi}\sqrt{-G}~{G}^{ZZ}
{G}^{x_1x_1}\left(\frac{\partial_{Z}E_{T}(Z)}{E_{T}(Z)}\right)\Biggr]_{Z_{\rm uv}},
\end{equation}
$\Omega_2$ being the volume of a two-sphere part of the world volume of the $D6$-branes. The coefficient of $E'_T(Z)/E_T(Z)$ was shown in \cite{Bulk2} to be rewritten as:
	\bg\label{jgaddar}
	e^{-\varphi}\sqrt{-G}~{G}^{ZZ}{G}^{x_1x_1} 
	 =  - 3 g_sN_f \left({g_sM^2\over N}\right)\left({\alpha^2_{\theta_1} \over \alpha^4_{\theta_2}}\right) \kappa_1 r_h^2
	\left(1 + \log~r_h\right), \nd
where $\kappa_1 \propto {1 \over g_s^{3/2}}$ with the constant of proportionality defined in eq. (6.66) of \cite{Bulk2}. Also, it was shown in \cite{Bulk2} that up to ${\cal O}\left(\frac{1}{N}\right)$:
\begin{equation}\label{elizshue}
{\bf Im}\left[{E_T'(Z)\over E_T(Z)}\right] = - {\omega \sqrt{4\pi g_sN}\over 4 r_h Z_{\rm uv}}\sqrt{6b^2 +1 \over 
	9b^2 + 1}, \end{equation}
where $b$ is the bare resolution parameter defined in \cite{Bulk2}.
Hence, the retarded Green's function in the zero momentum limit can be written as:
\begin{equation}\label{melleo}
\mathbb{G}_{x_1x_1}^{({\rm R})}(\omega, q = 0) \equiv 
\Omega_2 T_{D6}\left[e^{-\varphi}\sqrt{-G}~{G}^{ZZ}
{G}^{x_1x_1}\left(\frac{\partial_{Z}E_{T}(Z)}{E_{T}(Z)}\right)\right]_{Z_{\rm uv}}, \end{equation}
whose imaginary part yields the spectral function via: $\rho(T, \omega) \equiv -2 {\bf Im}~\mathbb{G}^{({\rm R})}(\omega, q = 0)$. It was shown in \cite{Bulk2} that up to ${\cal O}\left(\frac{1}{N}\right)$:
\begin{equation}\label{prunej}
{\rho(T, \omega)\over \omega} = {3\over 4}~g_sN_f \left({g_sM^2\over N}\right) \mathbb{F}_a(N, g_s, Z_{\rm uv}) 
\mathbb{F}_b(b, \alpha_{\theta_i}) ~a(r_h)~\log~r_h, 
\end{equation}
where $a(r_h)$, the full resolution parameter, is given in eq. (5.13) of \cite{Bulk2} and  $\mathbb{F}_a$ and $\mathbb{F}_b$ are now defined as:
\begin{equation}\label{tagchink}
\mathbb{F}_a(N, g_s, Z_{\rm uv}) = {N^{1/10}\sqrt{4\pi g_sN}\over g_s^{3/2} Z_{\rm uv}}, ~~~~~
\mathbb{F}_b(b, \alpha_{\theta_i}) = {\beta \over n_o} \left({\alpha^2_{\theta_1} \over \alpha^4_{\theta_2}}\right) 
\sqrt{6b^2 +1 \over 9b^2 + 1}. \end{equation}
If $\zeta_1 \equiv g_s, \zeta_2 \equiv 1/N$ and $\zeta_3 \equiv 1/Z_{\rm uv}$, then we can choose the behavior of each of these parameters such that:
\begin{equation}\label{teenmeye}
\lim_{\zeta_i \to 0} \mathbb{F}_a(\zeta_1, \zeta_2, \zeta_3) \equiv {\bf f}_a, \end{equation}     
with a constant ${\bf f}_a$ in the weak/strong string coupling and strong 't Hooft coupling limits. As $T \to 0$, $r_h$ vanishes, and hence $\lim_{\omega \rightarrow 0}\frac{\rho(T=0,\omega)}{\omega}$ also vanishes. Therefore:
\begin{equation}\label{jun17122}
\lim_{\omega \rightarrow 0}\frac{\rho(\omega)}{\omega} \equiv \lim_{\omega \rightarrow 0}\Biggl[\frac{\rho(T, \omega)}{\omega} - \frac{\rho(T=0, \omega)}{\omega}\Biggr]  \propto \frac{1}{3} - c_s^2,
\end{equation}
providing an inportant confirmation of our earlier analysis. 
We concluded \cite{Bulk2} with a discussion on the following. 

\vskip.1in

\noindent $\bullet$ As shown in \cite{NPB}, one can continue to work with classical supergravity despite working with, e.g., $g_s=0.45, M=3, N_f=2$ in the IR due to a color-flavor enhancement of the Planckian length scale.

\vskip.1in

\noindent $\bullet$ After cascading away all the $N$ $D3$-branes, using the NSVZ RG flow equation for the $SU(M)$ color gauge group that survives at the end of the Seiberg duality cascade, one can show that it indeed possible to obtain $g_{SU(M)}\sim{\cal O}(1)$ even if $g_s\rightarrow0$ in the strong 't Hooft coupling limit ($N\gg1: g_s N\gg1$) provided $N_f>0$.

\section{Conclusions and discussions}

In this paper we summarized some of the main results of \cite{Bulk2} associated with weak and strong 't Hooft couplings. The ratio of the bulk to shear viscosities is proportional to the square of the deviation of the sound speed from its conformal value at weak 't Hooft coupling, whereas at strong 't Hooft coupling the ratio is linearly proportional to the deviation. The behavior at intermediate  't Hooft coupling is unfortunately not tractable using available techniques, so the interpolating dynamics is presently unknown. Despite that progress can be made at the two extreme limits of the coupling range as shown in \cite{Bulk2}.

 \section*{Acknowledgements}

The work of AC is supported in part by the program Mobility Plus of the Polish Ministry of Science and Higher Education. The work of KD, CG and SJ is supported in part by the Natural Sciences and Engineering Research Council of Canada. CG gratefully acknowledges support from the Canada Council for the Arts through its Killam Research Fellowship program. 
The work of
KS is supported by a Senior Research Fellowship from the Ministry of Human Resource and Development, Government of India.
AM would like to thank McGill University for the wonderful hospitality during the completion of this work. AM is supported in part
by the Indian Institute of Technology, Roorkee, India and the department of Physics, McGill University, Canada.

\end{document}